\documentclass{ifacconf}

\usepackage{graphicx}      % include this line if your document contains figures
\usepackage{natbib}        % required for bibliography
\usepackage{epsfig}
\usepackage{graphicx}
\usepackage[normalsize,it,hang]{subfigure}
\usepackage{amsfonts}
\usepackage{amssymb,hhline}
\usepackage{graphics}
\usepackage{latexsym}
\usepackage{makeidx}
\usepackage{amsmath}
\usepackage{wasysym}
\usepackage{amscd}
\usepackage{color}
\usepackage{rotating}
\usepackage{array}

\newtheorem{remark}{Remark}[section]

\begin{document}
\begin{frontmatter}

\title{{\small{\tt Joint IFAC Conference: SSSC, TDS, COSY \\ Gif-sur-Vette, France, 30 June-2 July 2025}} \\ \vspace{2cm} Model-Free Predictive Control: \\ Introductory Algebraic Calculations, \\ and a Comparison with HEOL and ANNs}
%\thanksref{footnoteinfo}} 
%Title, preferably not more than 10 words.

%\thanks[footnoteinfo]{Sponsor and financial support acknowledgment
%goes here. Paper titles should be written in uppercase and lowercase
%letters, not all uppercase.}

\author[1,5]{Cédric Join}
\author[2]{Emmanuel Delaleau} 
\author[3,4,5]{Michel Fliess} 

\address[1]{CRAN (CNRS, UMR 7039), Universit\'{e} de Lorraine, BP 239, 54506 Vand{\oe}uvre-l\`{e}s-Nancy, France (e-mail: {cedric.join@univ-lorraine.fr})}
\address[2]{ENI Brest, UMR CNRS 6027, IRDL, 29200 Brest, France\\ (e-mail: emmanuel.delaleau@enib.fr)}
\address[3]{LIX (CNRS, UMR 7161), \'Ecole polytechnique, 91128 Palaiseau, France (e-mail: {michel.fliess@polytechnique.edu})}
\address[4]{LJLL (CNRS, UMR 7238), Sorbonne Unversit\'e, 75005 Paris, France (e-mail: {michel.fliess@sorbonne-universite.fr})}
\address[5]{AL.I.E.N., 7 rue Maurice Barr\`{e}s, 54330 V\'{e}zelise, France\\ (e-mail: \{{michel.fliess,cedric.join}\}{@alien-sas.com})}

\begin{abstract}  % Abstract of 50--100 words 
Model predictive control (MPC) is a popular control engineering practice, but requires a sound knowledge of the model. Model-free predictive control (MFPC), a burning issue today, also related to reinforcement learning (RL) in AI, is reformulated here via a linear differential equation with constant coefficients, thanks to a new perspective on optimal control combined with recent advances in the field of model-free control (MFC). It is replacing Dynamic Programming, the Hamilton-Jacobi-Bellman equation, and Pontryagin's Maximum Principle. The computing burden is low. The implementation is straightforward.
Two nonlinear examples, a chemical reactor and a two tank system, are illustrating our approach. A comparison with the HEOL setting, where some expertise of the process model is needed, shows only a slight superiority of the later. A recent identification of the two tank system via a complex ANN architecture might indicate that a full modeling and the corresponding machine learning mechanism are not always necessary neither in control, nor, more generally, in AI. 

\end{abstract}


Joint IFAC Conference: SSSC, TDS, COSY

\end{frontmatter}
%===============================================================================
\onecolumn
\section{Introduction}
Recent advances, with a strong algebraic flavor, in signal processing and time series analysis, have been applied to energy forecasting by \cite{solar} and \cite{energy}. Those forecasts are easier to implement than the current machine-learning approaches which are prominent now in AI. A similar goal is pursued here with {\em model-free predictive control} ({\em MFPC}), which is a natural extension of \emph{model-predictive control} ({\em MPC}), one of the most popular control engineering practice, where the mathematical modeling should be well known (see, e.g., \cite{rawlings}). Our approach to this hot topic (see, e.g., \cite{khal}, \cite{mesbah}) may be summarized as follows.
%\begin{enumerate}
%    \item 
    We use the {\em ultra-local model} of {\em model-free control} ({\em MFC}) (Fliess and Join (2013, 2022)) like \cite{belal}, \cite{feng}, \cite{he}, \cite{hegendus}, \cite{huo}, \cite{lammouchi}, \cite{li}, \cite{liu}, \cite{long}, \cite{mousavi}, \cite{rahmani},
\cite{sun1}, \cite{sun2}, \cite{sun3}, \cite{xu}, \cite{xu2}, \cite{wang}, \cite{yang}, \cite{yin}, \cite{yuan}, \cite{zhang1}, \cite{zhang2}, \cite{zhang3}, \cite{zhang4} \cite{zhou1}, \cite{zhou}.
%\item 
A new understanding of optimal control (\cite{join2,esaim}) via an algebraic interpretation of controllability (\cite{fliess90}, \cite{flat}) simplifies the connection between predictive and optimal controls.
%\begin{itemize}
%   \item 
    On a small time lapse, where a time-varying quantity may be approximated by a constant, the Euler-Lagrange equation, i.e., the fundamental equation of the calculus of variations (see, e.g., \cite{gelfand}), becomes a constant linear ordinary differential equation. 
%   \item 
    It leads to a dramatic simplification of the connection between predictive control and {\em reinforcement learning} (\emph{RL}), which is today a mainstay (see, e.g., \cite{recht}, \cite{adhau}, \cite{bertsekas}). It vindicates the viewpoint expressed by \cite{lecun}:  {\it I do favor MPC over RL. I've been making that point since at least 2016. RL requires ridiculously large numbers of trials to learn any new task}.
%\end{itemize}
%\end{enumerate}

If the key condition of a good model knowledge is not met, many studies use various artificial neural network (ANN) architectures to apply efficiently MPC techniques, without the need for MFPC. See, e.g., \cite{aa}, \cite{nubert}, \cite{ren}, \cite{salzmann}, \cite{adhau}, and references therein. The same two tank example used by \cite{adhau} via a \emph{recurrent neural network} (see, e.g., \cite{ag}) for its modeling is treated here with our ultra-local model of order $1$ (\cite{mfc1}), and yields, therefore, a trivial implementation. Is this an indication that ANNs do not always provide the best solution?
The other example studied is a chemical reactor (\cite{panno}, \cite{rawlings}), which was introduced in order to study the robustness of MPC with respect to disturbances. In those references, a time-invariant linearized modeling around a setpoint is used. It is therefore assumed that the reactor is always near the setpoint. This most stringent condition is lifted in our approach.
In both cases, we are comparing our results with the {\em HEOL} approach (\cite{heol}), already used by \cite{delaleau}, which is combining flatness-based and model-free controls. Although HEOL requires more process knowledge, its performance is not really superior.

%The above MFPC is compared with the HEOL setting recently introduced by \cite{heol}, %where an approximate (\emph{differentially}) \emph{flat} (\cite{flat}) modeling is %available. The linearization around an appropriate reference trajectory, which is %easily obtained thanks to the flatness property, yields a linear system, which is time-%varying in general. It leads to a \emph{homeostat}, which is analogous to the ultra-%local model for MFC. Local stability around trajectories and robustness are ensured by %an \emph{intelligent} controller, which was introduced by \cite{mfc1}. 

Our paper is organized as follows. MFPC is introduced in Sect.~\ref{mfpc} and HEOL in Sect.~\ref{setting}. Sect.~\ref{chem} (resp. \ref{tank}) depicts the chemical reactor (resp. the two tank system). See some concluding remarks in Sect.~\ref{conclu} and some reflections on modeling in AI. The mathematical modelings of the chemical reactor and of the two tank system are given in Appendices A and B for digital simulation purposes.

\section{Model-free predictive control}\label{mfpc}

\subsection{Ultra-local model}

With \cite{mfc1} consider for simplicity's sake the SISO (single-input single output) \emph{ultra-local model} or order $1$, which is replacing the poorly known plant and disturbance description  
\begin{equation}
\dot{y} = \mathcal{F} + \alpha u
\label{1}
\end{equation}
The control and output variables are respectively $u$ and $y$.
The derivation order of $y$ is $1$ like in many concrete situations.
$\mathcal{F}$ subsumes not only the unknown structure of the system, which most of the time is nonlinear, but also
any external disturbance. 
The constant $\alpha \in \mathbb{R}$ is chosen by the practitioner such that $\alpha u$ and $\dot{y}$ are of the same magnitude. Therefore $\alpha$ does not need to be precisely estimated.

A data-driven estimation of $\mathcal{F}$ in Eq. \eqref{1}, which is obtained via algebraic manipulations (\cite{garnier}), reads according to \cite{mfc1}:%\footnotesize
\begin{equation}\label{integral}
\begin{aligned}
 \mathcal{F}_{\text{est}}(t)  =-\frac{6}{T^3}\int_{0}^T& \left( (T -2\sigma)y(t-T+\sigma)\right.\\&\left.+\alpha(T-\sigma)\sigma u(t-T+\sigma) \right) d\sigma 
\end{aligned}
\end{equation}
\begin{remark}
Model-free control and the associated ultra-local model enjoy already many successes: See numerous references in Fliess and Join (2013, 2022), \cite{heol}, and, e.g., \cite{ait}, \cite{michel} for sensor fault accommodation and an airfoil benchmark).  
\end{remark}
\subsection{Optimization via the Euler-Lagrange equation}\label{calcul}
\subsubsection{Preliminary calculations.}
Assume that $F = a$ is a constant in Eq. \eqref{1}, which corresponds now to an elementary flat system, where $y$ is a flat output. Introduce the Lagrangian, or cost function, 
\begin{equation*}\label{lag}
\mathcal{L} = (y - y_{\rm setpoint})^2 + u^2 = (y - y_{\rm setpoint})^2 + \left(\frac{\dot{y} - a}{\alpha}\right)^2
\end{equation*}
where $y_{\rm setpoint}$ denotes a given setpoint. 
%Eq. \eqref{1} yields $$\mathcal{L} = (y - y_{\rm setpoint})^2 %+ \left(\frac{\dot{y} - a}{\alpha}\right)^2$$
For the criterion
$
    J = \int_{t_i}^{t_f} \mathcal{L}dt 
$
the Euler-Lagrange equation (see, e.g., \cite{gelfand}) 
$
\frac{\partial \mathcal{L}}{\partial y} - \frac{d}{dt} \frac{\partial \mathcal{L}}{\partial \dot{y}}  = 0
$
corresponds to a non-homogeneous linear ordinary diffrential equation of $2$nd order $\ddot{y} - \alpha^2 (y - y_{\rm setpoint}) = 0$.
%\begin{equation*}\label{ele}
%    \ddot{y} - \alpha^2 (y - y_{\rm setpoint}) = 0
%\end{equation*}
Any optimal solution  reads 
$%\begin{equation*}\label{sol}
    y^\star (t)= y_{\rm setpoint} + c_1\exp(\alpha t) + c_2\exp(-\alpha t)$, $c_1, c_2 \in \mathbb{R}$.
%\end{equation*}
It is independent of $a$; $c_1$, $c_2$ are obtained via the two-point boundary conditions $y(t_i) = y_i$, $y(t_f) = y_{\rm setpoint}$: $$c_1 =\frac{y_i\exp(-\alpha t_f) - y_{\rm setpoint}\exp(-\alpha t_f)}{\exp(\alpha t_i)\exp(-\alpha t_f) - \exp(-\alpha t_i)\exp(\alpha t_f)}$$
$$c_2 =-\frac{\exp(\alpha t_f)(y_i - y_{\rm setpoint})}{\exp(\alpha t_i)\exp(-\alpha t_f) - \exp(-\alpha t_i)\exp(\alpha t_f)}$$
%\begin{remark}\label{infinite}
%    If $\alpha > 0$, and $t_f = +\infty$ (infinite time horizon), Eq. \eqref{sol} becomes 
%$y^\star (t)= y_{\rm setpoint} + c_2\exp(- \alpha t)$.
%\end{remark}

\subsubsection{Application to MFPC.}
%\textcolor{red}{\texttt{[Previous version:]}
%Consider the time interval $0 \leqslant t \leqslant t_f$. %Subdivide it $0 < t_{\nu_\iota} \leq t_f$. At time %%$t_{\nu_\iota}$, replace $\mathcal{F}$ in Eq. \eqref{1} by the %constant $F_{\text{est}}(t_{\nu_\iota}) = a_{\nu_\iota}$. Reproduce the above optimization calculations on the horizon $t_{\nu_\iota} \leqslant t \leqslant t_f$. Iterate it on the interval $t_{\nu_\iota + 1} \leqslant t \leqslant t_f$.}
% \textcolor{green}{\texttt{[Suggestion:]}
% Subdivide the time interval $[t_i,t_f[$, $t_i< \cdots < t_k < t_{k+1} < \cdots < t_f$. On each time interval $[t_k,t_{k+1}[$ replace $\mathcal{F}$ of Eq.~\eqref{1} by a constant and reproduce the above optimization procedure on the time horizon $[t_{k+1},t_f[$.
% }
Consider the subdivision $0< \cdots < t_k < t_{k+1} < \cdots < t_f$. On each time lapse $[t_k,t_{k+1}]$ replace $\mathcal{F}$ in Eq.~\eqref{1} by the constant $\mathcal{F}_\text{est}(t_k)$. Start the above optimization procedure again on the time horizon $[t_{k+1},t_f]$. The criterion and the horizon may be modified if necessary.
%\begin{remark}
%It is difficult, if not impossible, to examine the stability without any further %assumptions on the domains where $y$ and $u$ should live.
%\end{remark}

\section{The HEOL setting}\label{setting}
\subsection{Tangent linear system}
%The calculations may be summarized as follows%\textcolor{red}{ici il faut un double indice %à $\nu_1$ et $\nu_2$}
Differentiate $\mathcal{A}_i$ in
$
\mathcal{A}_i (u_i, y_1, \dots, y_{1}^{(\nu_{i,1})}, y_2, \dots, y_{2}^{(\nu_{i,2})}) = 0$, $i = 1, 2$,
where $y_1$, $y_2$ (resp. $u_1$, $u_2$) are \emph{flat outputs} (resp. control variables) of a {\em flat} system (\cite{flat}):
\begin{equation*}\label{diff}
    \frac{\partial \mathcal{A}_i}{\partial u_i}du_i + \sum_{\iota = 1, 2}\frac{\partial \mathcal{A}_i}{\partial y_\iota} dy_\iota + \dots + \frac{\partial \mathcal{A}_i}{\partial y_{\iota}^{(\nu_\iota)}} d y_{\iota}^{(\nu_\iota)}  = 0
\end{equation*}
where $d y_{\iota}^{({\mu_\iota)}} = \frac{d^{\mu_\iota}}{dt^{\mu_\iota}} dy_\iota$. 
Associate to it the \emph{tangent}, or \emph{variational}, linear system, which is time-varying in general,
\begin{equation}\label{tang}
    \frac{\partial \mathcal{A}_i}{\partial u_i}\Delta u_i + \sum_{\iota = 1, 2}\left(\frac{\partial \mathcal{A}_i}{\partial y_\iota} + \dots + \frac{\partial \mathcal{A}_i}{\partial y_{\iota}^{(\nu_\iota)}} \frac{d^{\nu_\iota}}{dt^{\nu_\iota}}\right) \Delta y_{\iota}  = 0
\end{equation}
$\Delta u_\iota$ (resp. $\Delta y_\iota$) is a control (resp. output) variable.
\subsection{Homeostat}\label{homeos}
Assume that $\frac{\partial \mathcal{A}_\iota}{\partial \dot{y}_\iota} \neq 0$, $\iota = 1, 2$, in Eq. \eqref{tang}. Then write with \cite{heol} the \emph{homeostat}
\begin{equation}\label{hom}
    \frac{d}{dt} (\Delta y_\iota) = \mathfrak{F}_\iota + \alpha_\iota \Delta u_\iota, \quad \iota = 1, 2
\end{equation}
where
\begin{itemize}
    \item $\alpha_\iota = - \frac{\frac{\partial \mathcal{A}_i}{\partial u_i}}{\frac{\partial \mathcal{A}_i}{\partial \dot{y}_\iota}}$ is not constant in general;
    \item $\mathfrak{F}_\iota$ is not given by the missing terms in Eq. \eqref{tang}, but is data-driven, i.e., it corresponds to model mismatches and disturbances.
\end{itemize}
Write with \cite{heol} the following estimate $\mathfrak{F}_{\iota, \rm est}$, which is easily deduced from Eq. \eqref{integral}
{\small
\begin{equation}\label{integralbis}
\mathfrak{F}_{\iota, \rm est} = - \frac{6}{T^3} \int_{0}^{T} \left( (T - 2 \sigma)\Delta \tilde{y_\iota}(\sigma) {+} \sigma (T - \sigma)\tilde\alpha(\sigma)\Delta \tilde {u_\iota}(\sigma)\right)d\sigma
\end{equation}}
where 
\begin{itemize}
\item the time lapse $T > 0$  is ``small.''
\item $\Delta\tilde{y_\iota}(\sigma)=\Delta y_\iota(\sigma+t-T)$, $\tilde{\alpha}_\iota(\sigma)\Delta \tilde u(\sigma)= \alpha_\iota (\sigma+t-T)\Delta u_\iota (\sigma+t-T)$.
\end{itemize}
\subsection{iP controller}
Mimicking \cite{mfc1}, associate, with \cite{heol}, to the homeostat \eqref{hom} the \emph{intelligent proportional} (\emph{iP}) controller
\begin{equation}\label{ip}
 \Delta u_\iota = - \frac{\mathfrak{F}_{\iota, \rm est} + K_{\iota, P} \Delta y_\iota}{\alpha_\iota}, \quad \iota = 1, 2   
\end{equation}
where $K_{\iota, P} \in \mathbb{R}$ is the gain. Combine Eqs. \eqref{hom} and \eqref{ip}:
 $\left(\frac{d}{dt} + K_{\iota, P}\right) \Delta y_\iota  = 0$.
Thus $\lim_{t \to \infty} \Delta y_\iota \approx 0$ if, and only if, $K_{\iota, P} >0$.

\section{Computer experiments}\label{exp}
\subsection{Chemical reactor}\label{chem}
\subsubsection{Ultra-local model}
There are two control variables in the chemical reactor of Appendix \ref{A}. We are therefore using twice Eq. \eqref{1}. For the first (resp. second) case: $u = T_c$ (resp. $u = F$), $y = c$ (resp. $y = h$). 

\subsubsection{Homeostat for HEOL}
Eqs. \eqref{hom} and \eqref{chemical} yield%\footnote{See \cite{delaleau} for another %example.}

$$\frac{d}{dt} {\Delta c} = \mathfrak{F}_c - \frac{2{EUk_0}\exp(-\frac{E}{RT})c}{RC_pr\rho T^2} \Delta u_c$$
and 
$$\frac{d}{dt} {\Delta h} = \mathfrak{F}_h - \frac{1}{\pi r^2} \Delta u_h$$

\subsubsection{Simulations}
\begin{figure*}[!ht]
\centering%
\subfigure[\footnotesize $T_c$ (-- blue) and nominal control (- - red)]
{\epsfig{figure=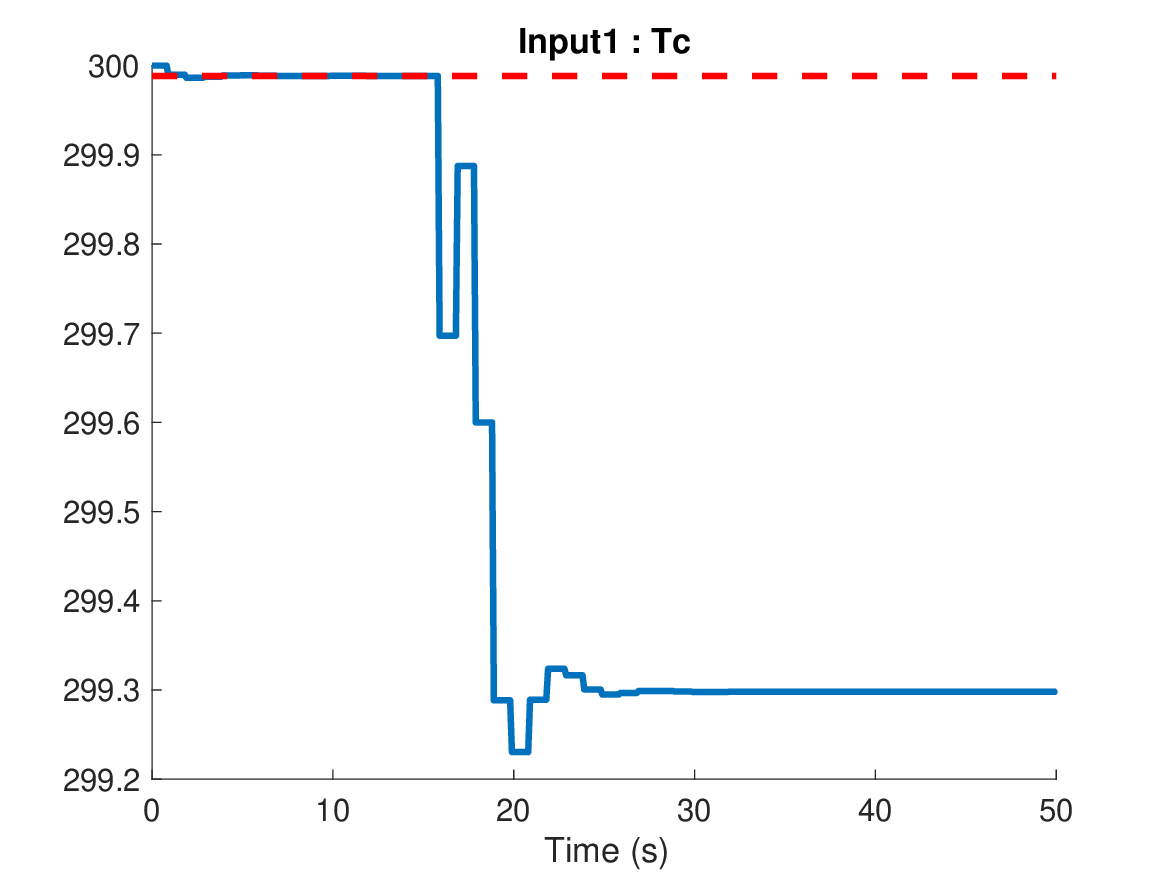,width=0.31\textwidth}}%\hspace{.5cm}
\subfigure[\footnotesize $F$ (-- blue) and nominal control (- - red)]
{\epsfig{figure=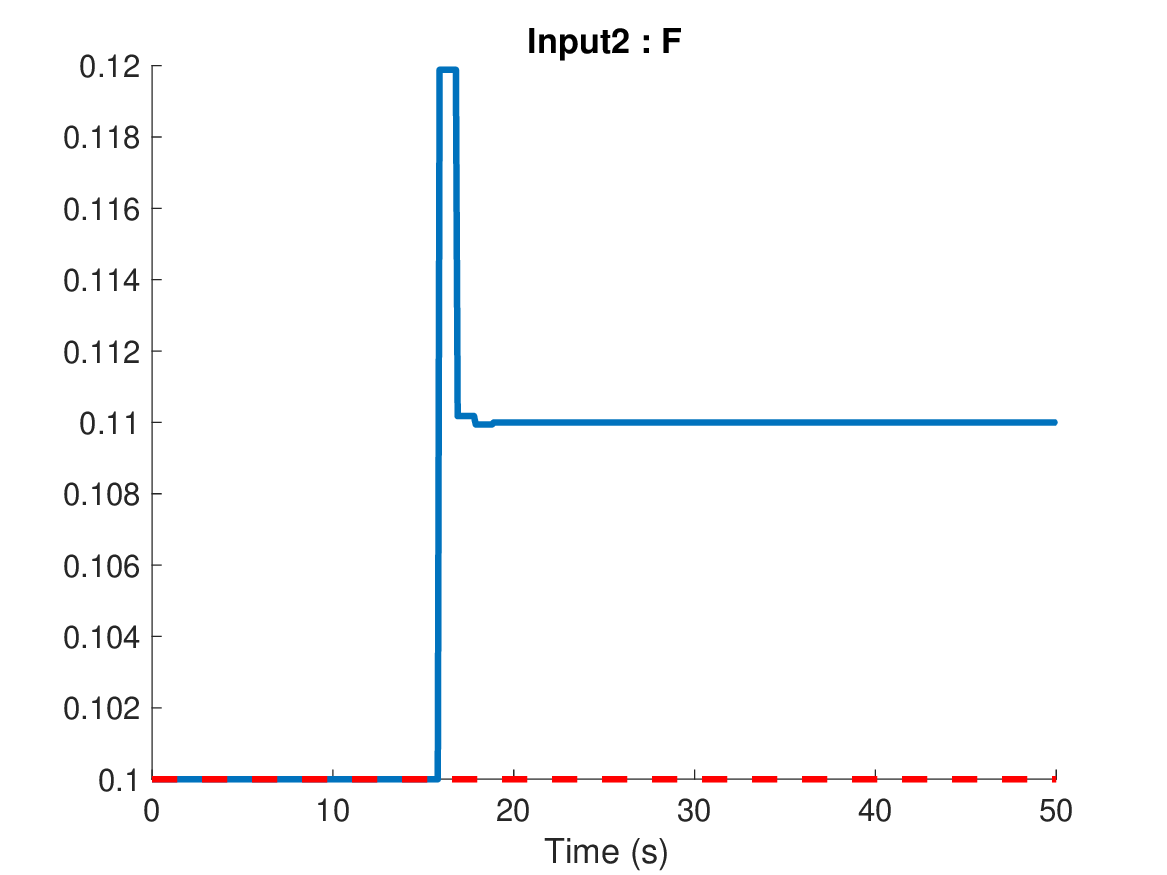,width=0.31\textwidth}}%\hspace{.5cm}
\\
\subfigure[\footnotesize $c$ (-- black) and its reference trajectory (- - red)]
{\epsfig{figure=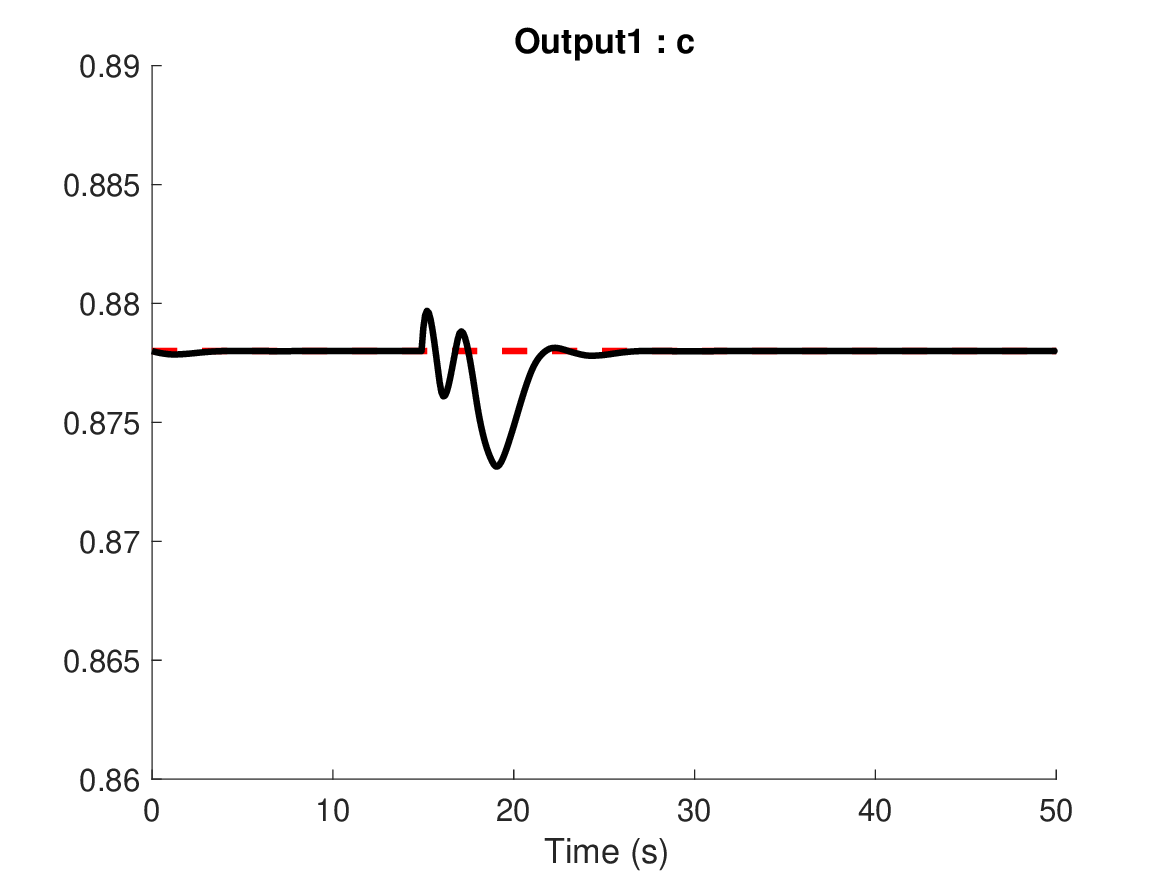,width=0.31\textwidth}}%\hspace{.5cm}
\subfigure[\footnotesize $T$]
{\epsfig{figure=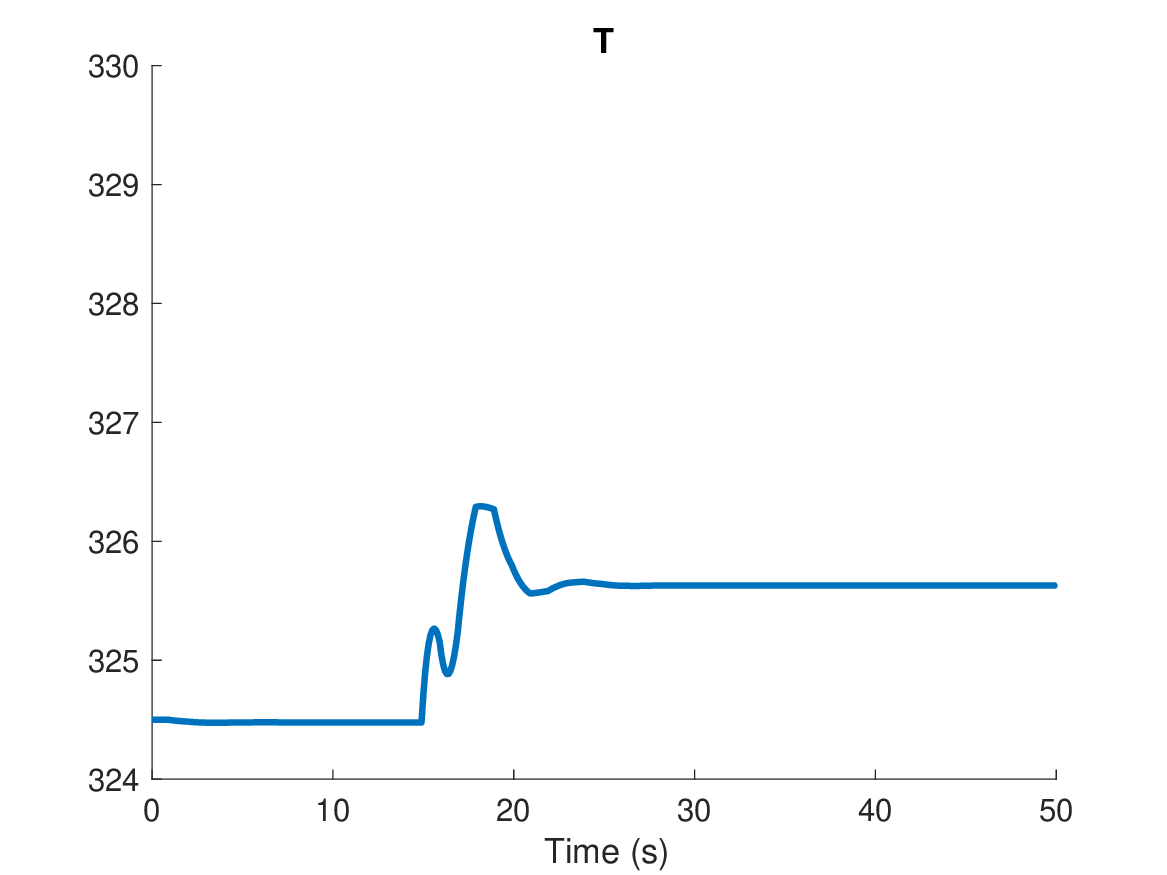,width=0.31\textwidth}}%\hspace{.5cm}
\subfigure[\footnotesize $h$ (-- black) and its reference trajectory (- - red)]
{\epsfig{figure=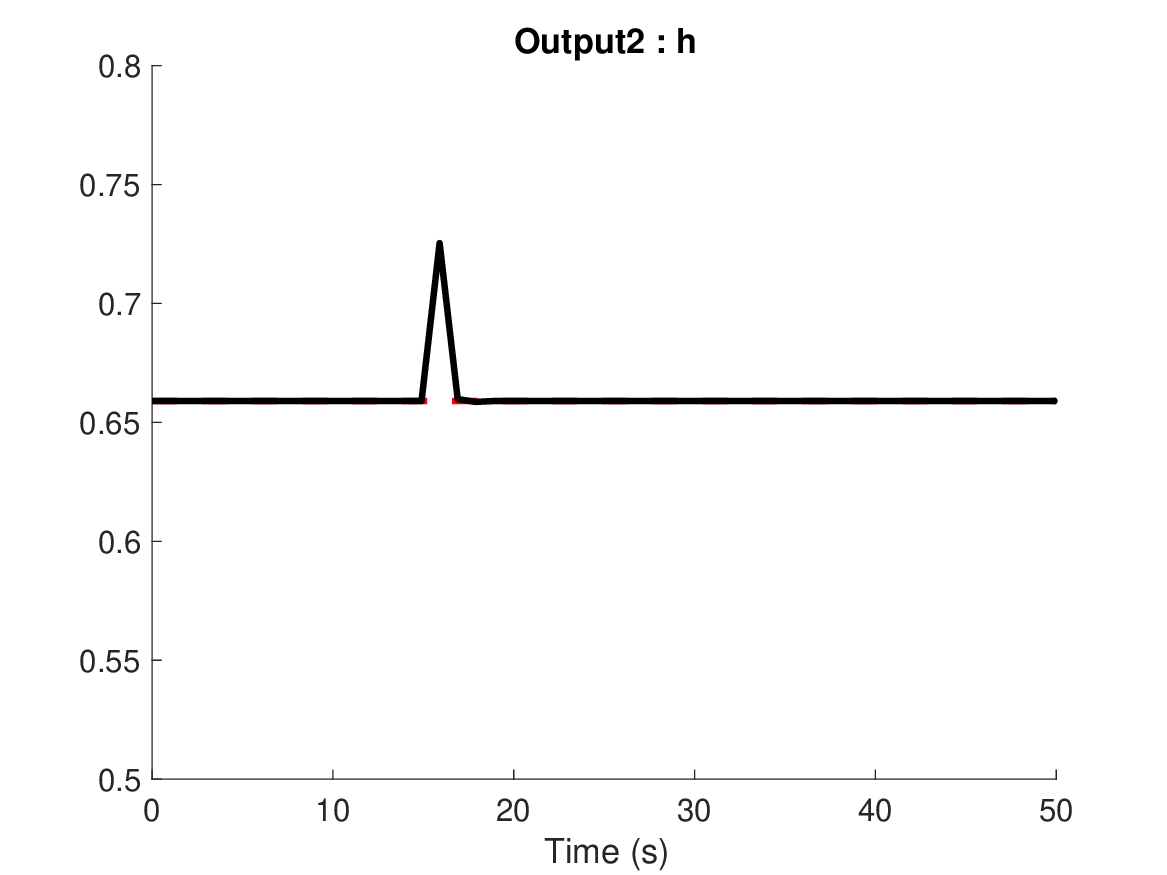,width=0.31\textwidth}}%\hspace{.5cm}
\caption{Chemical reactor: HEOL}\label{H2}
\end{figure*}

\begin{figure*}[!ht]
\centering%
\subfigure[\footnotesize $T_c$]
{\epsfig{figure=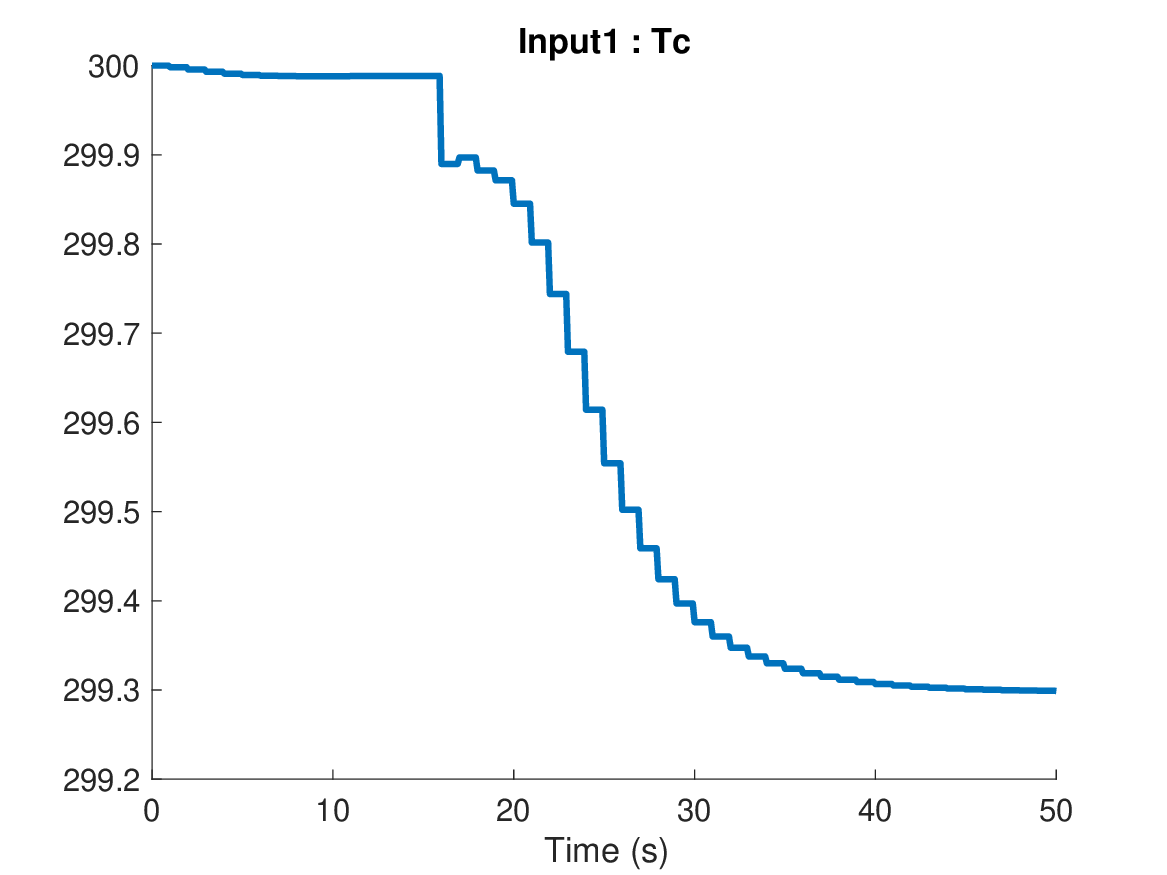,width=0.31\textwidth}}%\hspace{.5cm}
\subfigure[\footnotesize $F$]
{\epsfig{figure=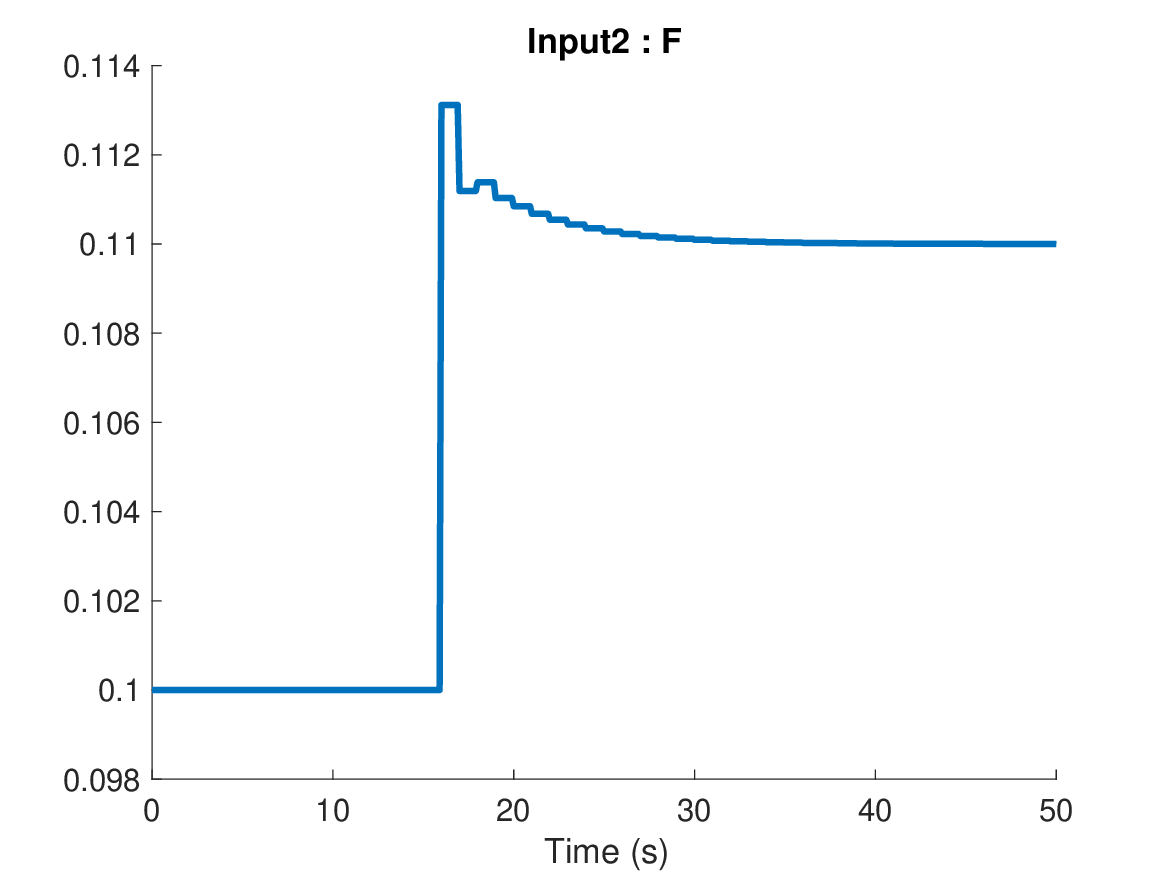,width=0.31\textwidth}}%\hspace{.5cm}
\\
\subfigure[\footnotesize $c$ (-- black) and its set-point (- - red)]
{\epsfig{figure=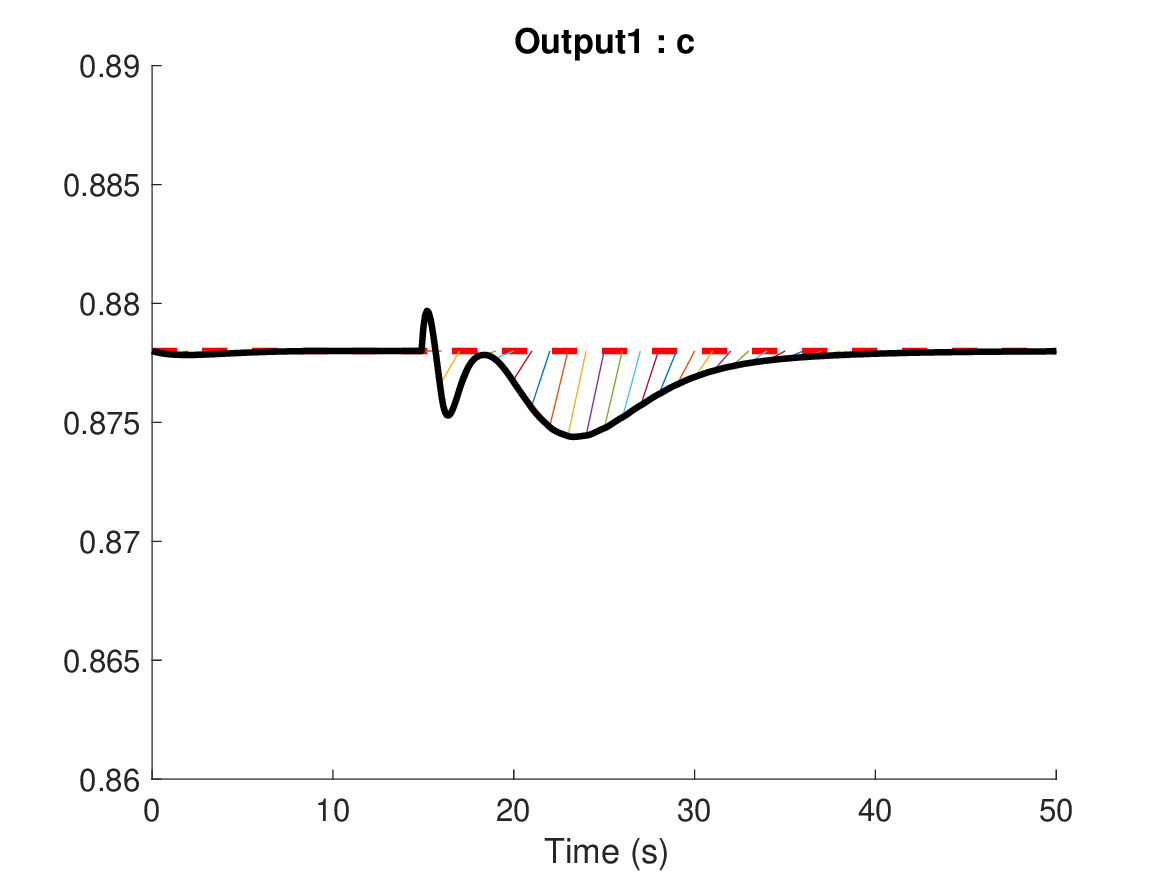,width=0.31\textwidth}}%\hspace{.5cm}
\subfigure[\footnotesize $T$]
{\epsfig{figure=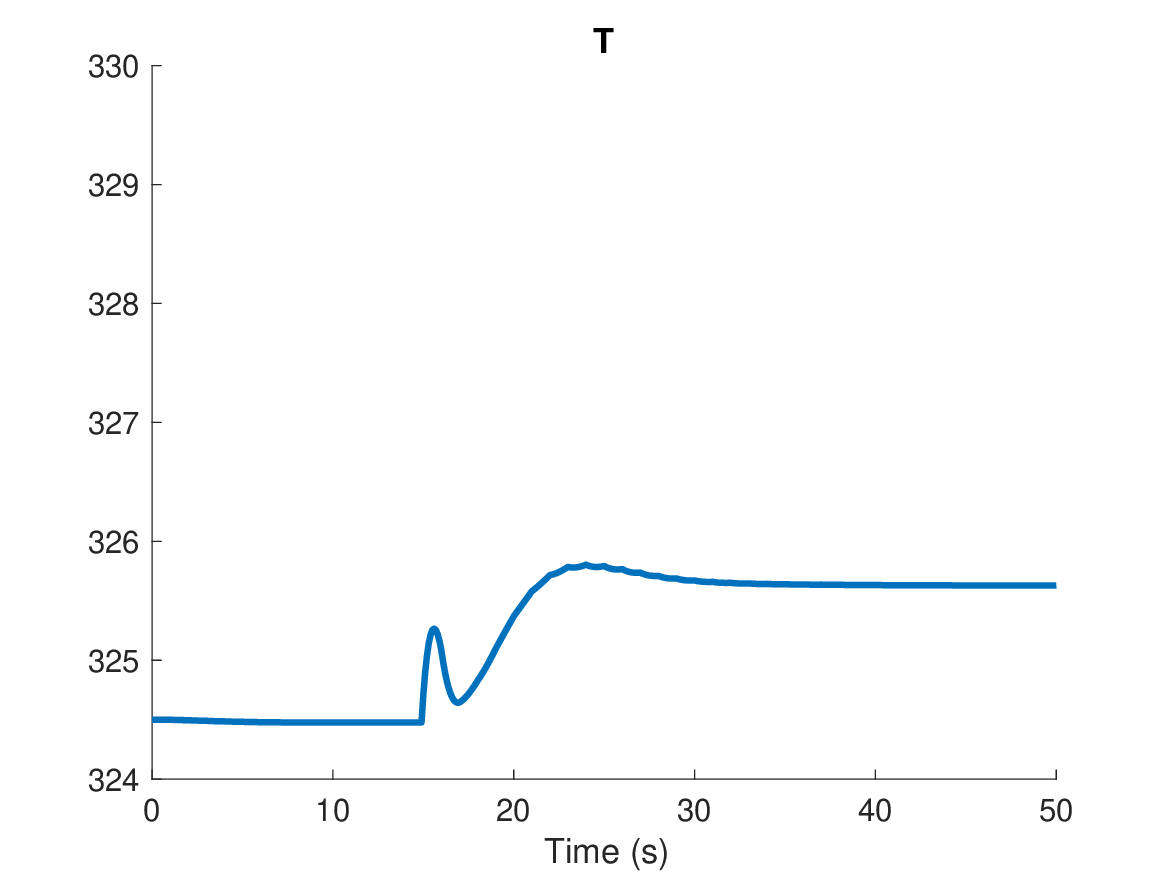,width=0.31\textwidth}}%\hspace{.5cm}
\subfigure[\footnotesize $h$ (-- black) and its set-point (- - red)]
{\epsfig{figure=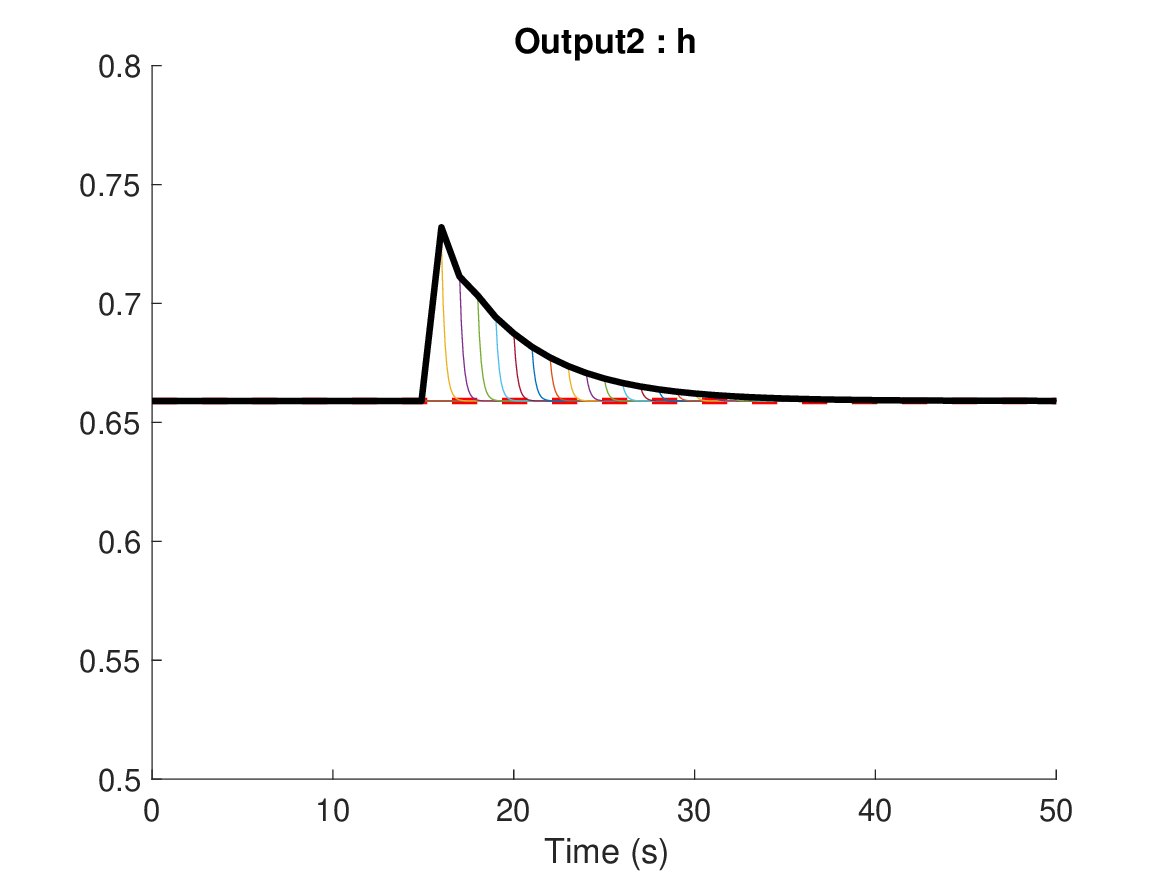,width=0.31\textwidth}}%\hspace{.5cm}
\caption{Chemical reactor: MFPC}\label{M2}
\end{figure*}

We follow \cite{panno} and \cite{rawlings}. For the initial operating point, $c(0) = 0.878$\,kmol/m$^3$, $T(0) = 324.5$\,K, $h(0) = 0.659$\,m, $T_c(0) = 300$\,K, $F(0) = 0.1$\,m$^3$/min. The sampling time for the control variables (resp. Eq. \eqref{integral} and \eqref{integralbis}) is $1$\,min (resp. $0.1$\,min). The total duration is $50$\,min. For MFPC, set for the first (resp. second) ultra-local model \eqref{1} $\alpha = -0.05$ (resp. $\alpha = - 5$). According to Sect. \ref{calcul}, a new optimal trajectory is calculated every minute. For HEOL, set $K = 1$ for the gain in the iP \eqref{ip}. 
Reference trajectories are deduced from classic B\'{e}zier curves (see, \textit{e.g.}, \cite{rogers}). Introduce the following disturbance: An unmeasured increase of the flow F of $10\,\%$. We seek to maintain the initial operating point for $c$ and $h$: See Figs. \ref{H2} and \ref{M2}.%, and \ref{I2}.
The figures above show a certain superiority of HEOL over MFPC. It should be added, however, that MFPC's results are satisfactory and, above all, do not require any knowledge of the model.
 
% une élévation du débit $F$ non mesuré de $10\%$ perturbe le bon fonctionnement et nous %cherchons à conserver le point de fonctionnement initial pour $c$ et $h$ (voir Figures %\ref{H2}, \ref{M2} et \ref{I2}).

% \begin{figure*}[!ht]
% \centering%
% \subfigure[\footnotesize $T_c$ (-- blue) and nominal control (- - red)]
% {\epsfig{figure=H1Tc.eps,width=0.214\textwidth}}%\hspace{.5cm}
% %
% \subfigure[\footnotesize $F$ (-- blue) and nominal control (- - red)]
% {\epsfig{figure=H1F.eps,width=0.214\textwidth}}%\hspace{.5cm}
% %
% \\
% \subfigure[\footnotesize $c$ (-- black) and its reference trajectory (- - red)]
% {\epsfig{figure=H1c.eps,width=0.214\textwidth}}%\hspace{.5cm}
% %
% \subfigure[\footnotesize $T$]
% {\epsfig{figure=H1T.eps,width=0.214\textwidth}}%\hspace{.5cm}
% %
% \subfigure[\footnotesize $h$ (-- black) and its reference trajectory (- - red)]
% {\epsfig{figure=H1h.eps,width=0.214\textwidth}}%\hspace{.5cm}
% %
% \caption{HEOL: setpoint change}\label{H1}
% \end{figure*}
%%
%%%%%%
% \begin{figure*}[!ht]
% \centering%
% \subfigure[\footnotesize $T_c$]
% {\epsfig{figure=M1Tc.eps,width=0.214\textwidth}}%\hspace{.5cm}
% %
% \subfigure[\footnotesize $F$]
% {\epsfig{figure=M1F.eps,width=0.214\textwidth}}%\hspace{.5cm}
% %
% \\
% \subfigure[\footnotesize $c$ (-- black) and its set-point (- - red)]
% {\epsfig{figure=M1c.eps,width=0.214\textwidth}}%\hspace{.5cm}
% %
% \subfigure[\footnotesize $T$]
% {\epsfig{figure=M1T.eps,width=0.214\textwidth}}%\hspace{.5cm}
% %
% \subfigure[\footnotesize $h$ (-- black) and its set-point (- - red)]
% {\epsfig{figure=M1h.eps,width=0.214\textwidth}}%\hspace{.5cm}
% %
% \caption{MFPC: setpoint change}\label{M1}
% \end{figure*}
%%
%%%%%
% \begin{figure*}[!ht]
% \centering%
% \subfigure[\footnotesize Input 1]
% {\epsfig{figure=I1Tc.eps,width=0.45\textwidth}}%\hspace{.5cm}
% %
% \subfigure[\footnotesize Input 2]
% {\epsfig{figure=I1F.eps,width=0.45\textwidth}}%\hspace{.5cm}
% %
% \\
% \subfigure[\footnotesize Measured output 1]
% {\epsfig{figure=I1c.eps,width=0.214\textwidth}}%\hspace{.5cm}
% %
% \subfigure[\footnotesize Output 2]
% {\epsfig{figure=I1T.eps,width=0.214\textwidth}}%\hspace{.5cm}
% %
% \subfigure[\footnotesize Measured output 3]
% {\epsfig{figure=I1h.eps,width=0.214\textwidth}}%\hspace{.5cm}
% %
% \caption{MFPC$\infty$ : setpoint change}\label{I1}
% \end{figure*}
% %%
% \begin{figure*}[!ht]
% \centering%
% \subfigure[\footnotesize Input 1]
% {\epsfig{figure=I2Tc.eps,width=0.45\textwidth}}%\hspace{.5cm}
% %
% \subfigure[\footnotesize Input 2]
% {\epsfig{figure=I2F.eps,width=0.45\textwidth}}%\hspace{.5cm}
% %
% \\
% \subfigure[\footnotesize Measured output 1]
% {\epsfig{figure=I2c.eps,width=0.214\textwidth}}%\hspace{.5cm}
% %
% \subfigure[\footnotesize Output 2]
% {\epsfig{figure=I2T.eps,width=0.214\textwidth}}%\hspace{.5cm}
% %
% \subfigure[\footnotesize Measured output 3]
% {\epsfig{figure=I2h.eps,width=0.214\textwidth}}%\hspace{.5cm}
% %
% \caption{MFPC$\infty$ : perturbation rejection}\label{I2}
% \end{figure*}

\subsection{Two tank system}\label{tank}
\subsubsection{Ultra-local system}
Use Eq. \eqref{1}, where, according to Appendix B, $y = h_2$.

%The system consists of two tanks with cross-sections $s_1$ and $s_2$. A controlled %inlet flow $u>0$ feeds the first tank. The free outlet flow $q_{12}>0$ feeds the %second tank, which also has a free outlet flow $q_2>0$.

%Le système est constitué de deux cuves de section $s_1$ et $s_2$. Un débit commandé %d'entrée $u>0$ alimente la première cuve. Le débit de sortie, libre, noté $q_{12}>0$ %alimente la seconde cuve qui possède également un débit de sortie, libre, noté %$q_2>0$.\\
%En utilisant les équations de conservation, on écrit 
%\begin{equation}
%\begin{cases}
%s_1 \frac{d}{dt} h_1=u-q_{12}\\
%s_2 \frac{d}{dt} h_2= q_{12}-q_2
%\end{cases}
%\end{equation}

%où $10>h_1>0$ et $10>h_2>0$ sont les hauteurs d'eau dans les cuves.\\
%Il est habituel de modéliser un débit de sortie libre en utilisant les lois de %Bernoulli soit $q_{12}=k_1'\sqrt{h_1}$ et  $q_{2}=k_3'\sqrt{h_2}$ où $k_1'$ et $k_2'$ %sont liés à la viscosité du fluide. Ceci permet alors l'écriture 
%\begin{equation}\label{model1}
%\begin{cases}
%s_1 \dot h_1=u-k_1'\sqrt{h_1}\\
%s_2 \dot h_2= k_1'\sqrt{h_1}-k_3'\sqrt{h_2}
%\end{cases}
%\end{equation}
%le mesure $y=h_2$ est également sortie plate, notée $y^\star$.\\
%En divisant par $s_1$ et $s_2$ les deux équations différentielles du modèle %\eqref{model1} et en posant $k_1=k_1'/s_1$, $k_4=1/s_1$, $k_2=k_1'/s_2$ et %$k_3=k_3'/s_2$ on retrouve les constantes du benchmark "Cascaded Tanks System with %Overflow" décrit https://www.nonlinearbenchmark.org.

\subsubsection{Homeostat}
Eq. \eqref{bernoulli} yields by differentiation:
$$d \dot h_2=\frac{1}{s_2}\left( -\frac{k_2}{2\sqrt{h_2}}dh_2-s_1d \dot h_1+d u \right)$$
%où $\dot h_1$ est indépendant de $du$.\\
The corresponding homeostat reads :
$$\frac{d}{dt} {\Delta h_2} = \mathfrak{F} + \frac{1}{s_2} \Delta u$$

Thus $\alpha=\frac{1}{s_2}$ is constant.

% Avec la sortie plate $y^\star$ et donc la trajectoire de référence $h_2^\star$, on détermine 
% $$h_1^\star=\left(\frac{s_2 \dot y^\star+k_3'\sqrt{y^\star}}{k_1'} \right)^2$$
% et 
% $$u^\star=s_1\dot h_1^\star+k_1'\sqrt{h_1^\star}$$
% La somme des deux équations différentielles donne 
% $$s_1 \dot h_1+s_2 \dot h_2=u-k_3'\sqrt{h_2}$$
% soit
% $$s_2 \dot h_2=u-k_3'\sqrt{h_2}-s_1 \dot h_1$$
% Il est alors évident que $\alpha=\frac{1}{s_2}$ est constant et que $\nu=1$ en référence à la forme du modèle ultra-local \eqref{1}. Cela mène à $\mathcal{C}(\Delta y)= K_p \Delta y$.

%Pour MFPC, on applique scrupuleusement la démarche expliquée plus haut en privilégiant %la solution économe.

\subsubsection{Simulations}
\begin{figure*}[!ht]
\centering
\subfigure[\footnotesize $u^\star$ (- -) and $u$ (--)]
{\epsfig{figure=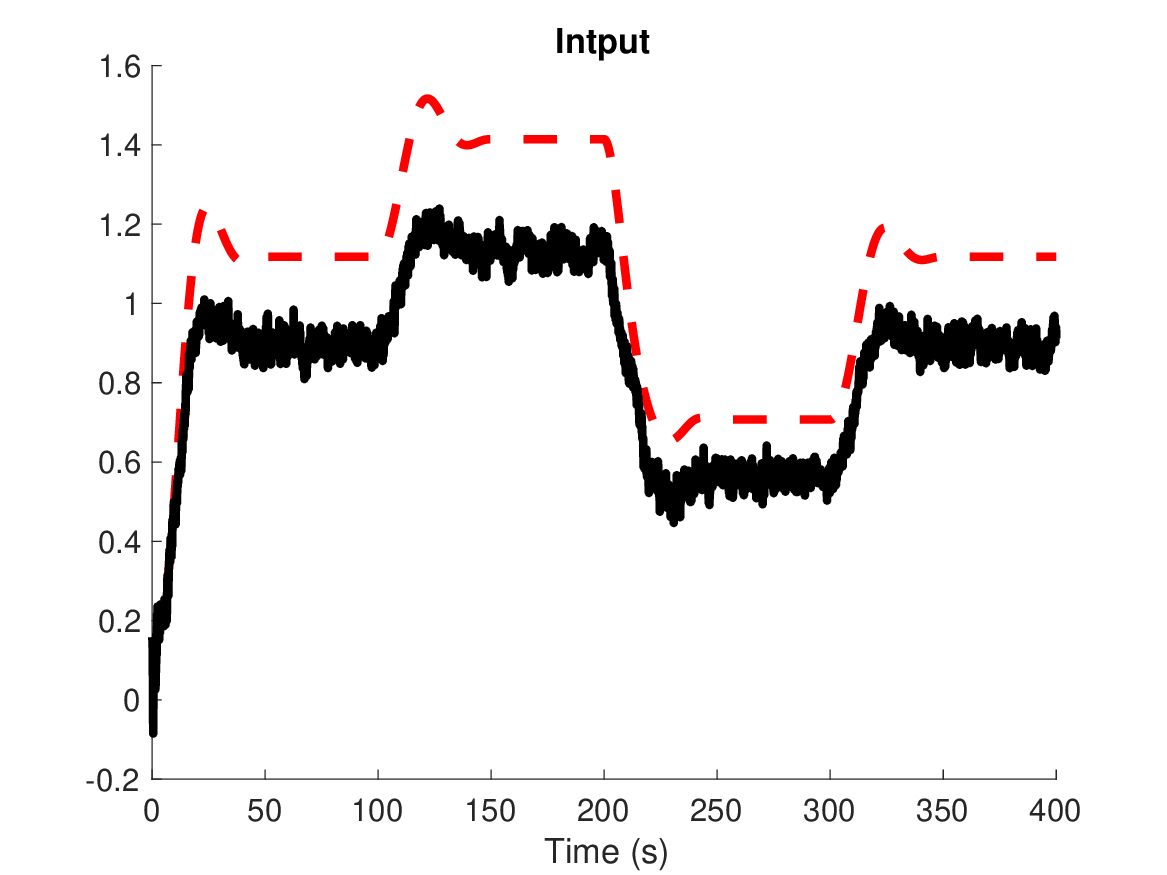,width=0.31\textwidth}}%\hspace{.5cm}
\subfigure[\footnotesize $y^\star$ (- -) and $y$ (--)]
{\epsfig{figure=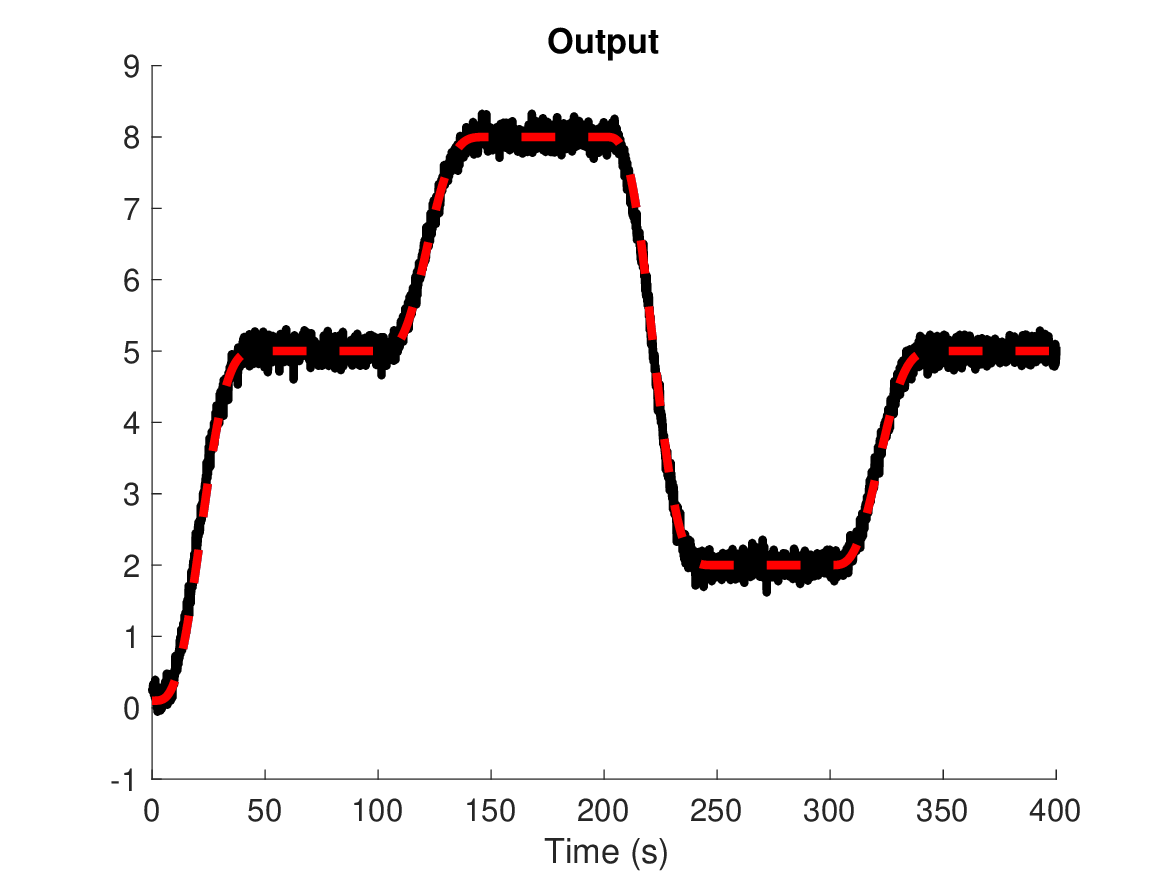,width=0.31\textwidth}}%\hspace{.5cm}
\subfigure[\footnotesize Tank levels $h_1$  and $h_2$]
{\epsfig{figure=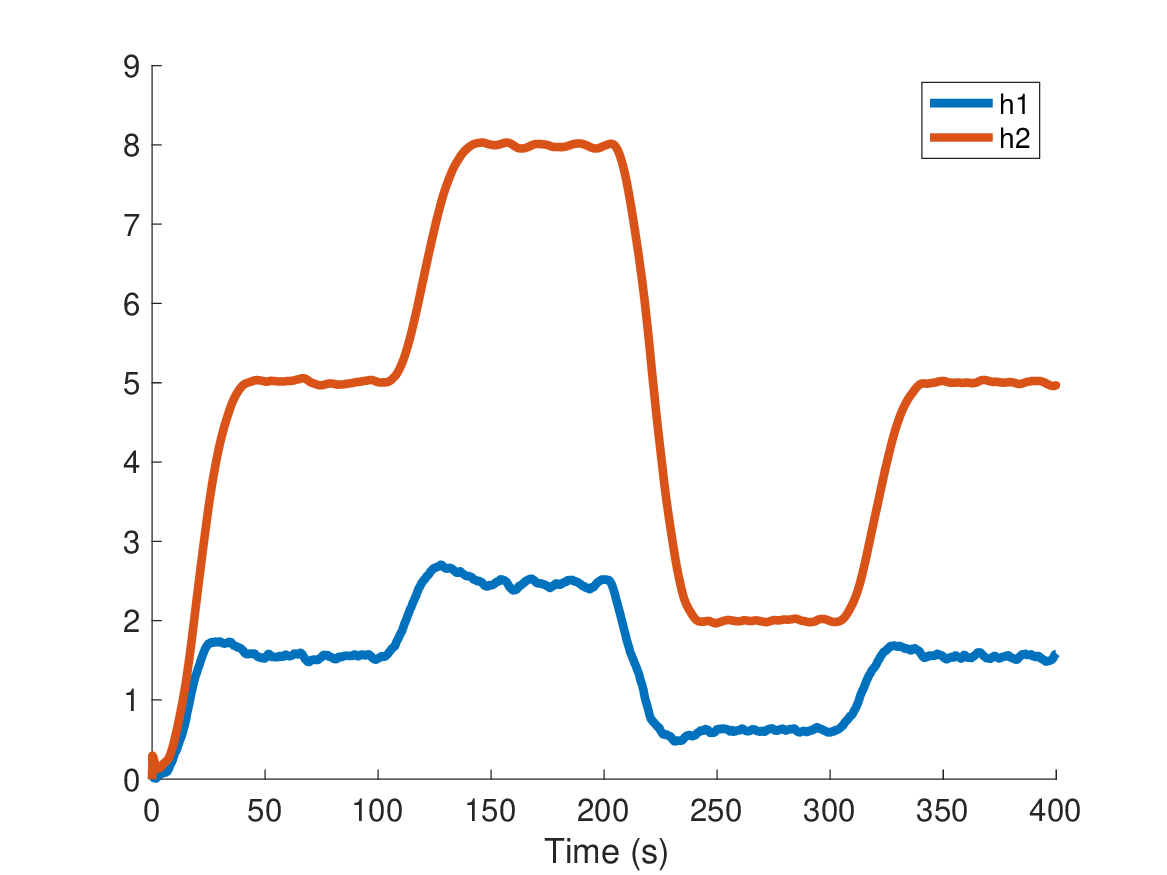,width=0.31\textwidth}}%\hspace{.5cm}
\caption{Two tanks: HEOL}\label{HU}
\end{figure*}

An additive white Gaussian noise $\mathcal{N}(0,0.1)$ corrupts the measurement.
Simulations last $400$s with a sampling period of $0.1$s.
For the iP controller associated to the homeostat command set $K=0.1$. For the MFPC the time horizon is $2$s. Our experiments mimic \cite{adhau}.
Introduce an uncertainty $\nabla=0.2$: The parameters are now $k_1(1+\nabla)$ and $k_2(1-\nabla)$. Without changing the control law parameters, the tracking remains excellent. As shown in Fig. \ref{HU}, the control variable differs nevertheless from the nominal one. According to Fig. \ref{MU}, the  setpoints are well reached via the MFPC despite the uncertainty. Moreover the constraints in Appendix B on the control variable and the water levels are respected.

% Un bruit additif ($\mathcal{N}(0,0.1)$) corrompt la mesure.
%Les simulations durent $400s$ avec une période d'échantillonnage de $0.1s$.
%Pour la commande HEOL, nous fixons $K=0.1$ et pour la MFPC, $T=2s$.
%\\
%Nous avons reproduit la même évolution du setpoint que celle de l'article [Gros 24].\\
% En l'absence d'incertitude HEOL corrige légèrement la commande nominale pour faire face à l'erreur initiale de $0.2$ sur la mesure. Comme le montre la Figure \ref{HN}, la poursuite de trajectoire est parfaite avec $u^\ast$ (resp. $y^\ast$) et $u$ (resp. $y$) pratiquement confondues.\\\\
%L'introduction d'une incertitude $\nabla=0.2$ modifie les paramètres $k_1(1+\nabla)$ %et $k_2(1-\nabla)$. Sans changer le paramètrage de la loi de commande, la poursuite %reste parfaite. Comme on peut le remarquer, Figure \ref{HU}, la commande appliquée %diffère de $u^\ast$ calculée sur le modèle nominale.\\
%\\
%En utilisant ce modèle incertain, la MFPC est évaluée, Figure \ref{MU}. Les Setpoint %sont atteints sans défaut et sans aucune connaissance modèle.

% \begin{figure*}[!ht]
% \centering
% \subfigure[\footnotesize $u^\star$ (- -) and $u$ (--)]
% {\epsfig{figure=HU.eps,width=0.214\textwidth}}%\hspace{.5cm}
% %
% \subfigure[\footnotesize $y^\star$ (- -) and $y$ (--)]
% {\epsfig{figure=HY.eps,width=0.214\textwidth}}%\hspace{.5cm}
% %
% \subfigure[\footnotesize Tank levels $h_1$  and $h_2$]
% {\epsfig{figure=HH.eps,width=0.214\textwidth}}%\hspace{.5cm}
% %
% \caption{HEOL : nominal case}\label{HN}
% \end{figure*}

\begin{figure*}[!ht]
\centering
\subfigure[\footnotesize $u$ (--)]
{\epsfig{figure=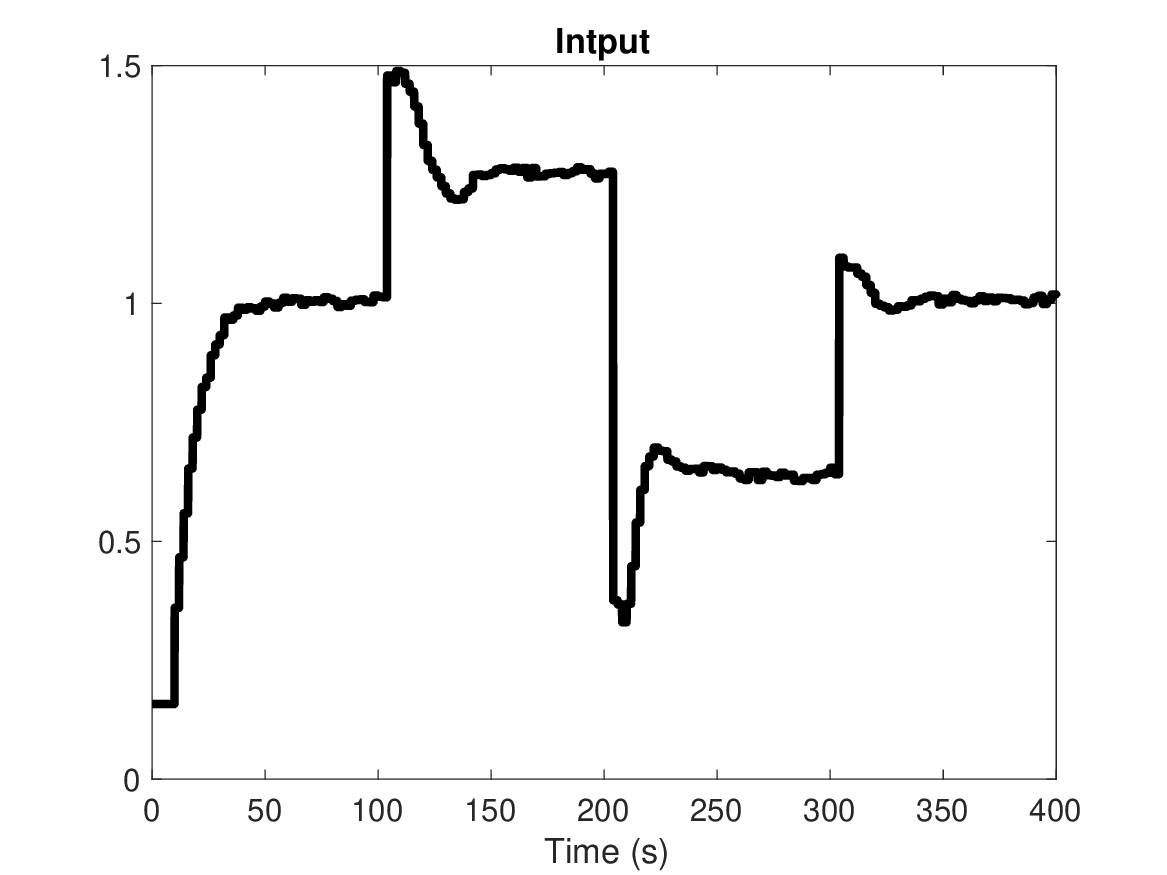,width=0.31\textwidth}}%\hspace{.5cm}
\subfigure[\footnotesize Setpoint (- -) and $y$ (--)]
{\epsfig{figure=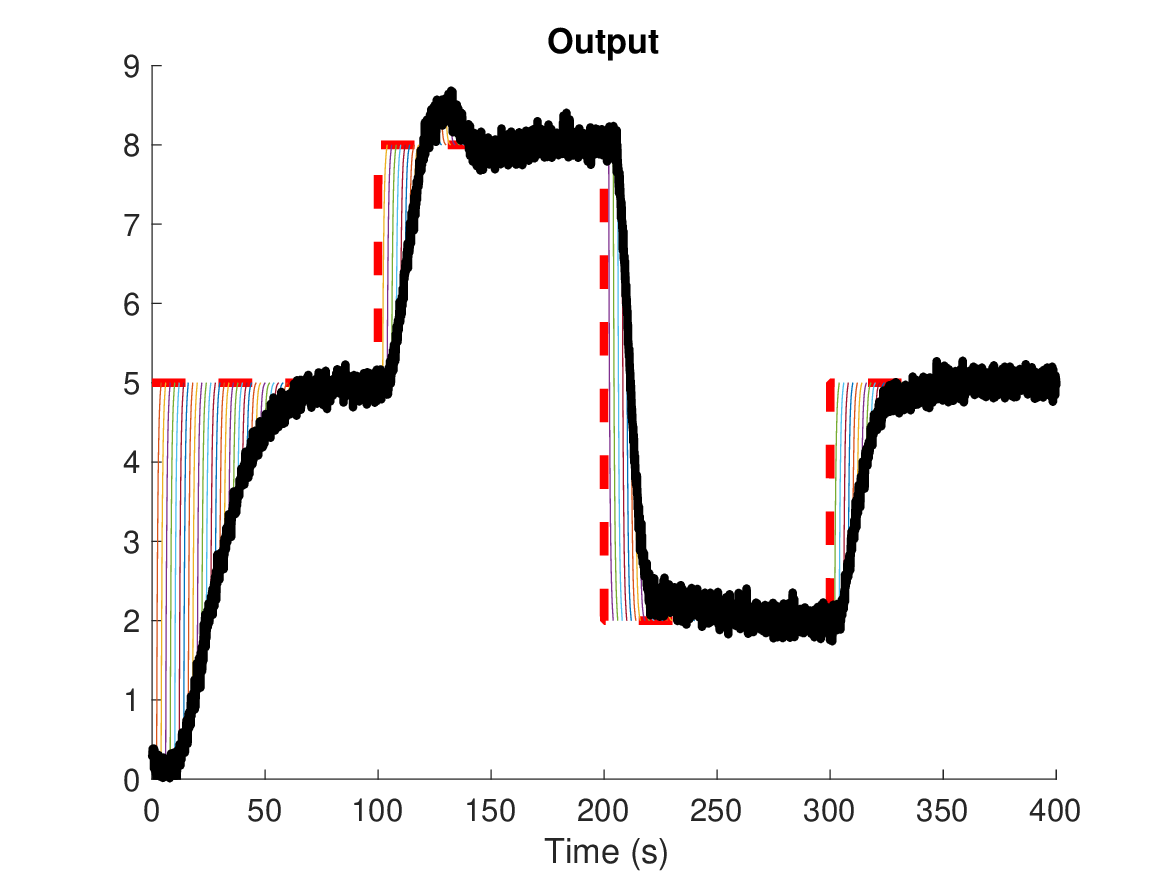,width=0.31\textwidth}}%\hspace{.5cm}
\subfigure[\footnotesize Tank levels $h_1$  and $h_2$]
{\epsfig{figure=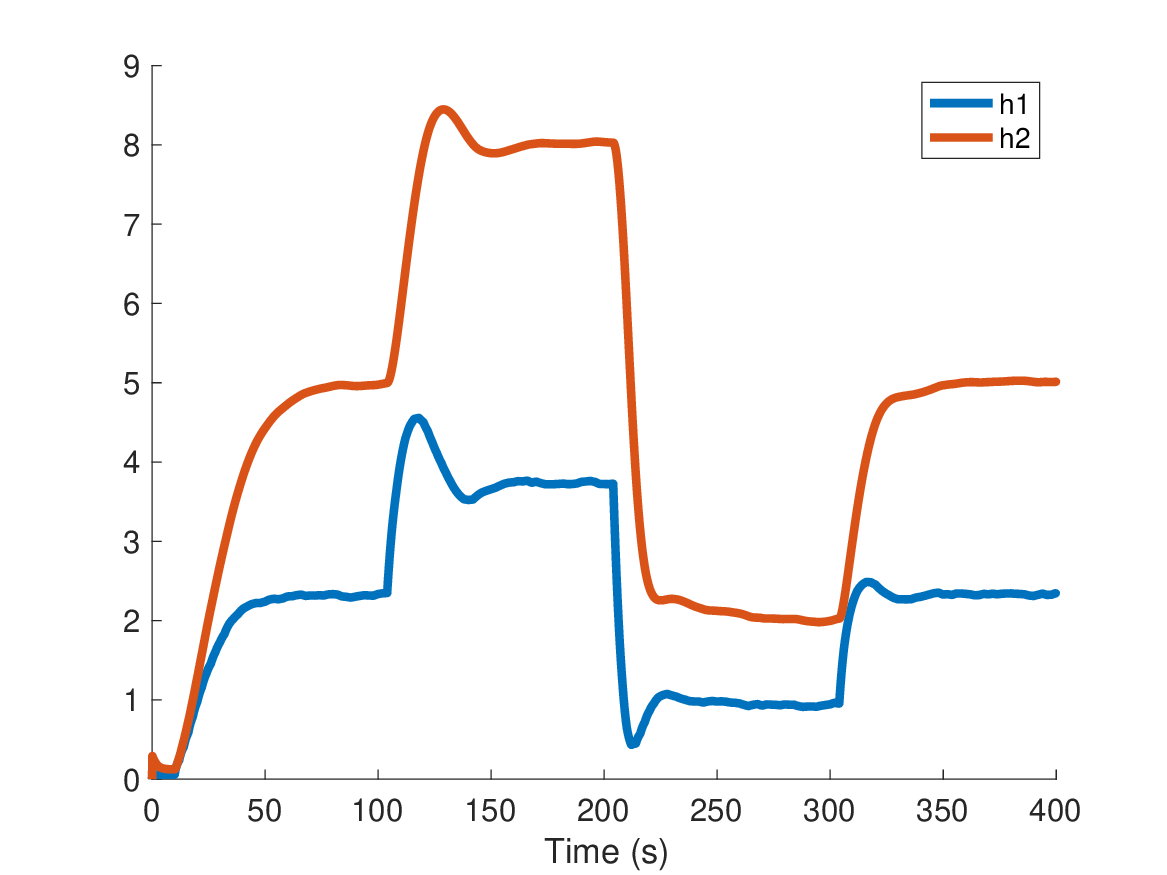,width=0.31\textwidth}}%\hspace{.5cm}
\caption{Two tanks: MFPC}\label{MU}
\end{figure*}

\cleardoublepage 
\section{Conclusion}\label{conclu}
Although the above simulations look encouraging, most important issues on constraints and stability are missing. Promising preliminary results will soon be presented.

% The stunning successes of modern AI is often based on models as complete as possible and on the associated machine-learning mechanism. The now classic terminology {\em large language model} ({\em LLM}) bears witness to this, as well as to the colossal energy resources required for its development (see, e.g., \cite{llm}). The results reported here may cast some doubt on the universality of this point of view, and, perhaps, pave the way for less greedy techniques.

% \textcolor{red}{The remarkable advances in modern AI often rely on models striving for comprehensiveness, along with their associated machine learning mechanisms. The term 'large language model' (LLM) exemplifies this trend towards comprehensive modeling, reflecting the significant energy resources required for their development (see, e.g., \cite{llm}). The results presented here suggest a complementary perspective to this view, and may open the way for exploring more resource-efficient techniques.}

% \textcolor{green}{
The remarkable advances in modern AI often rely on models striving for comprehensiveness, along with their associated machine learning mechanisms. While outside the realm of control theory, the term {\em large language model} ({\em LLM}) exemplifies this trend towards comprehensive modeling, reflecting the significant energy resources required for their development (see, e.g., \cite{llm}). \cite{suck} has said more than once about LLM and generative AI: \textit{Machine learning sucks}. This  skepticism is not alien to our techniques, which might be more subtle, at least in some parts of control engineering and robotics.

%Let us emphasize once again the strong algebraic flavor of our tools:
%flatness-based control (\cite{flat}), MFC (Fliess and Join (2013, 2022)), HEOL %%%    %(\cite{heol}), optimal control (\cite{join2}). 

%Most of our tools have an algebraic origin:
%\begin{itemize}
%    \item The Euler-Lagrange equation in optimal control was introduced (\cite{join2}) %via the module-theoretic interpretation of linear controllability (\cite{fliess90}).
%    \item Key parts of MFC have a deep algebraic flavor, for instance Formulae %\eqref{integral} and \eqref{integralbis} for data-driven estimations of key quantities %by employing algebraic estimation techniques (\cite{garnier}). 
%    \item Flatness (\cite{flat}) and the HEOL setting (\cite{heol}) were defined via %the language of differential algebra. 
%\end{itemize}

%\begin{ack}
%Place acknowledgments here.
%\end{ack}

%\bibliography{ifacconf}             % bib file to produce the bibliography
 % with bibtex (preferred)
%\clearpage  

\newpage

\appendix
\section{A chemical reactor}\label{A}  
\begin{figure}[!ht]
\centering
\subfigure[\footnotesize Well-stirred reactor]
{\epsfig{figure=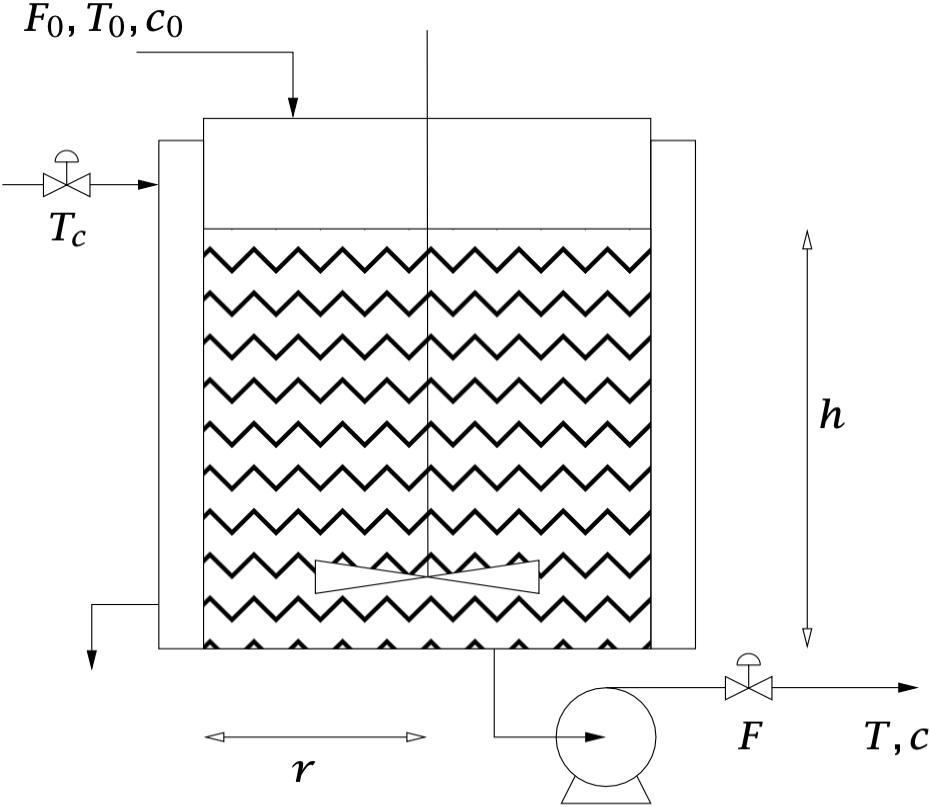,width=0.4\textwidth}}%\hspace{.5cm}
\subfigure[\footnotesize Two tanks]
{\epsfig{figure=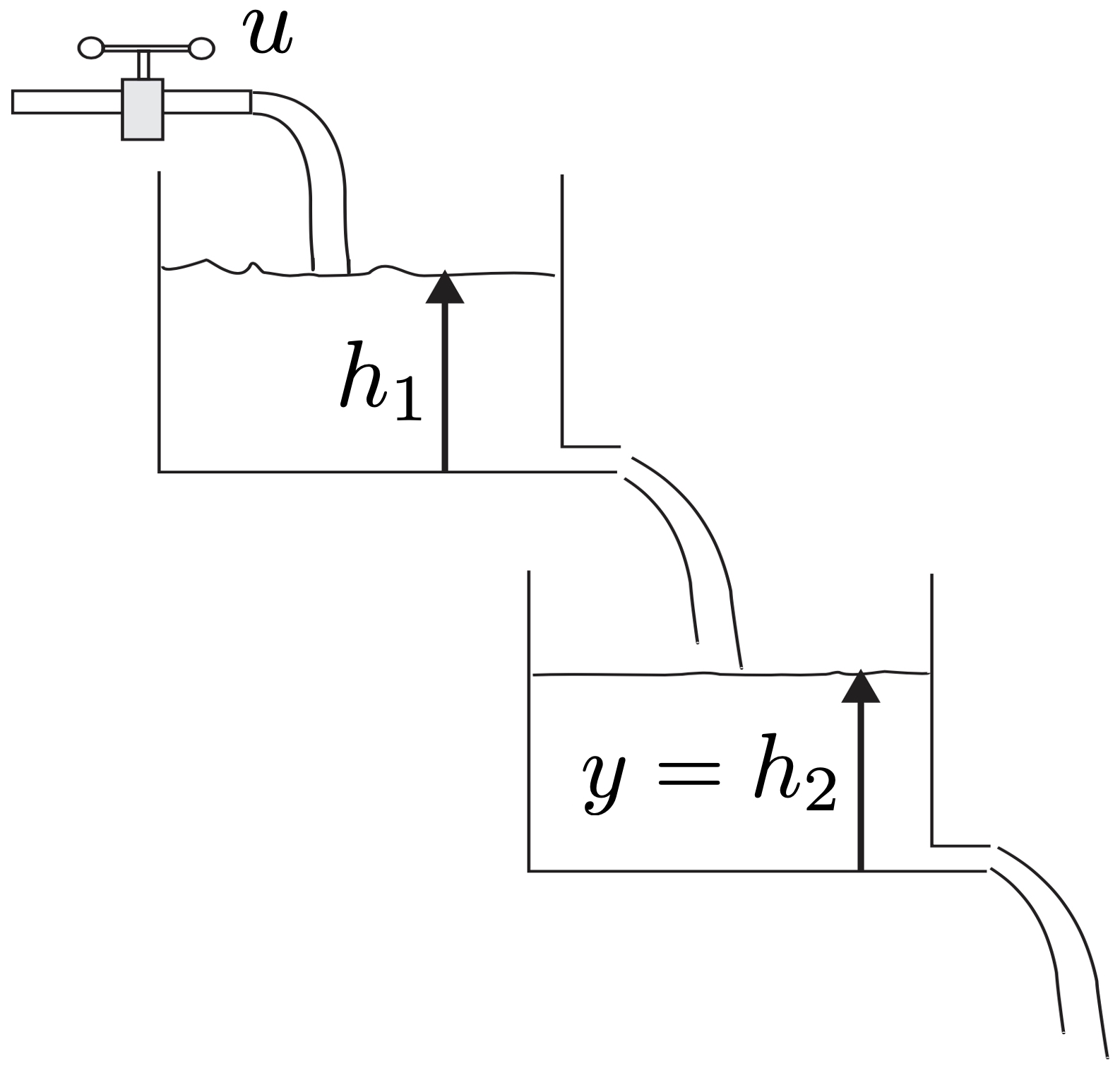,width=0.4\textwidth}}%\hspace{.5cm}
\caption{Process schemes}\label{Process}
\end{figure}

%\subsection{System description}
Fig. \ref{Process}-(a) depicts a well-stirred chemical reactor, which was introduced by \cite{panno} and \cite{rawlings} for investigating the robustness of linear time-invariant MPC with respect to disturbances. The following nonlinear model is used:
{\tiny 
\begin{equation}\label{chemical}
\begin{cases}
 \dot{c}=\frac{F_0(c_0-c)}{\pi r^2h}-k_0\exp{\left(-\frac{E}{RT}\right)}c\\
  \dot T=\frac{F_0(T_0-T)}{\pi r^2h}-\frac{\Delta H}{\rho C_p}k_0\exp{\left(-\frac{E}{RT}\right)}c+\frac{2U}{r\rho C_p}(T_c-T)\\
  \dot h=\frac{F_0-F}{\pi r^2}
  \end{cases}
\end{equation}} 
\noindent where $c$ (resp. $h$) is the fluid concentration (resp. height). Both quantities are measured. The control variables are $F$ and $T_c$. %\footnote{In the above references, a linearized version of Eq. \eqref{chemical},around an equilibrium point, is employed.} 
In the computer experiments $F_0=0.1$, $T_0=350$, $c_0=1$, $r=0.2149$, $k_0=7.2.10^{10}$, $\frac{E}{R}=8750$, $U=54.94$, $\rho=1000$, $C_p=0.2149$, $\Delta H=-5.10^4$.

%\begin{figure}[!ht]
%\centering
%\subfigure[\footnotesize Well-stirred reactor]
%{\epsfig{figure=React.jpg,width=0.214\textwidth}}%\hspace{.5cm}
%
%\subfigure[\footnotesize Two tanks]
%{\epsfig{figure=2tank,width=0.214\textwidth}}%\hspace{.5cm}
%
%\caption{Process schemes}\label{Process}
%\end{figure}

% \begin{figure}[!ht]
% \centering%
% %\subfigure[\footnotesize Measured output 2 (-- blue) and reference trajectory (- - red)]
% {\epsfig{figure=React.jpg,width=0.214\textwidth}}%\hspace{.5cm}
% %
% \caption{Well-stirred reactor}\label{React}
% \end{figure}

%\subsection{Flatness}\label{appflat}
Eq. \eqref{chemical} defines a \emph{flat} system (\cite{flat}). It is straightforward to check that $c$ and $h$ are \emph{flat outputs}: %All other system variables may be %expressed as \emph{differential functions} of $c$ and $h$, i.e., as functions of $c$, %$h$ and their derivatives up to some finite order: 
$F=-(\dot h\pi r^2-F_0)$, and $T_c=\frac{(\dot T-F_0\frac{T_0-T}{\pi r^2 h}+\frac{\Delta H}{\rho C_p}k_0\exp(-\frac{E}{R T})c)(r\rho C_p)}{2U}+T$, where $T=\frac{-R}{E \log(\dot c+F_0 \frac{c_0-c}{\pi r^2 h k_0 c})}$.

\section{A two tank system}\label{B} 
Consider with \cite{adhau} the two tanks system described by \cite{schoukens}:\footnote{Fig. \ref{Process}-(b) is borrowed from \cite{easy}, where this system is examined via algebraic estimation techniques.}
{\tiny 
\begin{equation}\label{bernoulli}
\begin{cases}
s_1\dot{h}_1=u-k_1\sqrt{h_1}\\
s_2\dot{h}_2= k_1\sqrt{h_1}-k_2\sqrt{h_2}
\end{cases}
\end{equation}}
\noindent where $u \geq 0$ is the control variable, $h_\iota$, $\iota = 1, 2$, $0 \leq h_\iota \leq 10$, is the water level, $k_\iota$ and $s_\iota$ are constant parameters. Eq. \eqref{bernoulli} defines a flat system, where $h_2$ is a flat output. 
%We obtain
%$$h_1=\left(\frac{s_2 \dot h_2+k_2\sqrt{h_2}}{k_1} \right)^2$$
%and 
%$$u=s_1\dot h_1+k_1\sqrt{h_1}$$
In the computer experiments $s_1, s_2 =1$, $k_1=0.6$, $k_2=0.5$.

%\section{The HEOL setting}              %Sections and %subsections %are supported  
                                                                         % in the appendices.
\end{document}